\documentclass[floatfix,prd,twocolumn,letterpaper,lengthcheck,superscriptaddress,showpacs,amssymb,amsmath,amsfonts,aps,altaffilletter,nofootinbib,nopreprintnumbers,showpacs]{revtex4-1}

\usepackage{color}
\usepackage{hyperref}
\usepackage{amsmath}
\usepackage{multirow}
\usepackage{graphicx}

\newcommand{\gwparams}{\ensuremath{\vec{\vartheta}}}
\newcommand{\ip}[2]{\left< #1, #2 \right>}
\newcommand{\kombine}{\texttt{kombine}}
\newcommand{\tref}{\ensuremath{t_{\mathrm{ref}}}}
\newcommand{\ttaper}{\ensuremath{t_{\mathrm{taper}}}}
\newcommand{\tgrid}{\ensuremath{t_{\mathrm{grid}}}}
\newcommand{\thmeco}{\ensuremath{t_{\mathrm{hMECO}}}}

\begin{document}

\title[]{Observational tests of the black hole area increase law}

\author{Miriam Cabero}
\email{miriam.cabero@aei.mpg.de}
\author{Collin D. Capano} 
\author{Ofek Fischer-Birnholtz}
\author{Badri Krishnan}
\author{Alex B. Nielsen}
\author{Alexander H. Nitz}
\affiliation{Max Planck Institute for Gravitational Physics (Albert Einstein Institute), Callinstrasse 38, D-30167 Hannover, Germany}
\affiliation{Leibniz Universit\"at Hannover, Welfengarten 1-A, D-30167 Hannover, Germany}
\author{Christopher M. Biwer}
\affiliation{Department of Physics, Syracuse University, Syracuse, NY 13244, USA}
\affiliation{Applied Computer Science (CCS-7), Los Alamos National Laboratory, Los Alamos, NM 87545, USA}

\date{\today}

\begin{abstract}

  The black hole area theorem implies that when two black holes merge,
  the area of the final black hole should be greater than the sum of
  the areas of the two original black holes. We examine how this
  prediction can be tested with gravitational-wave observations of
  binary black holes. By separately fitting the early inspiral and
  final ringdown stages, we calculate the posterior distributions for
  the masses and spins of the two initial and the final black holes.
  This yields posterior distributions for the change in the area and
  thus a statistical test of the validity of the area increase law.  We
  illustrate this method with a GW150914-like binary black hole
  waveform calculated using numerical relativity, and detector
  sensitivities representative of both the first observing run and
  the design configuration of Advanced LIGO. We obtain a $\sim74.6\%$
  probability that the simulated signal is consistent with the area
  theorem with current
  sensitivity, improving to $\sim99.9\%$ when Advanced LIGO reaches
  design sensitivity.  An important ingredient in our test is a method
  of estimating when the post-merger signal is well-fit by a damped
  sinusoid ringdown waveform.

\end{abstract}

\maketitle

\section{Introduction}
\label{sec:intro}

The black hole area increase law~\cite{Hawking:1971tu, Hawking1972} is one of the most celebrated results in exact non-linear relativity. It applies to both single black holes and to mergers of multiple black holes and implies that the total horizon areas should always increase provided certain assumptions that are expected to hold for astrophysical black holes. This law
suggests an analogy between the area of the horizon of a black hole and its
entropy \cite{Bekenstein:1972tm, Bardeen:1973gs, Bekenstein:1973ur}, and is thus one of
the central results of black hole thermodynamics.  The observations of
gravitational waves from binary black hole coalescence events
\cite{Abbott:2016blz,Abbott:2016nmj,TheLIGOScientific:2016pea, PhysRevLett.118.221101, GW170814} open up the possibility of
carrying out direct tests of this fundamental law using observational data.
In this paper we shall present a method to perform such a test
based on the inspiral and on the ringdown stages of simulated binary black hole coalescences.

Gravitational-wave observations of binary black hole coalescences correspond to the inspiral of two black holes that merge together to form a single black hole which then settles down to a stationary state by a ringdown process. The areas of the black hole horizons are not directly observable using gravitational wave data, so here we make use of the Kerr metric to relate the black hole horizon area to its asymptotic mass $M$ and dimensionless spin $\chi$. Widely separated inspiralling black holes are expected to be very well described by the Kerr metric in some neighbourhood of the horizon. Mass and spin values can be inferred for the inspiralling black holes by comparison with inspiral gravitational waveforms. The mass and spin of the final black hole can be inferred by comparison with ringdown waveforms of a single Kerr black hole.

If two initially distant Kerr black holes with areas $A_1, A_2$ coalesce to form a final Kerr black hole with area $A_f$, then it follows from the
area increase law that
\begin{equation}
\label{eq:a1a2}
  A_1 + A_2 \equiv A_i < A_f\,.
\end{equation}

Our analysis strategy is similar to the suggestion by Hughes and Menou \cite{Hughes:2004vw}, namely, use the early inspiral regime to
measure the parameters of the initial black holes, and independently use the late-time ringdown regime
to measure the parameters of the final black hole.  The end result of the
parameter estimation procedure will be a probability distribution
$p(\Delta A)$ for the change in the area
$\Delta A := A_f - (A_1 + A_2)$.  This allows us to calculate a
probability
\begin{equation}
\int_0^\infty p(\Delta A) d\Delta A  
\end{equation}
that a particular event is compatible with the area increase law.

By analysing the observed gravitational wave data for the detected events
~\cite{Abbott:2016blz, Abbott:2016nmj, TheLIGOScientific:2016pea, PhysRevLett.118.221101, GW170814, GW170817},
posterior distributions for the parameters of
the initial and final black holes in these events have been reported
in \cite{TheLIGOScientific:2016wfe,TheLIGOScientific:2016pea, PhysRevLett.118.221101, GW170814}.
However, these cannot be used as an independent test of
the area increase law because these results use fitting formulae to obtain
the final $M_f,\chi_f$ (and thus $A_f$) as functions of the initial
parameters. These fitting formulae~\cite{Healy:2014yta, Hofmann2016, Jimenez-Forteza:2016oae}
are results of numerical relativity simulations
which assume the validity of vacuum general relativity, 
and thus implicitly assume \emph{apriori} the validity of the area increase law.
A true direct test of the area increase law should independently
obtain $A_i$ and $A_f$ from the observed data and then use them to verify whether
or not Eq.(\ref{eq:a1a2}) holds.

Similarly, tests have been designed to check the consistency between
the inspiral and merger-ringdown parts of gravitational wave signals
\cite{Ghosh:2016qgn, ghosh2017testing}. When performed on gravitational
wave observations these tests have found the signals to be compatible
with standard general relativity and the fitting formulae
\cite{Abbott:2016blz, TheLIGOScientific:2016pea,
  PhysRevLett.118.221101, GW170814}. Our proposed test here
is different in that we avoid the use of fitting formulae entirely; we
separate the signal in the time domain rather than the frequency
domain, and crucially we avoid using data from the near merger portion
of the signal. A violation of the area theorem is perhaps most likely
near the merger of the two black holes where the spacetime is most
dynamic. It is this region where numerical relativity simulations are
critical to follow the evolution. We wish to avoid this region and
infer areas for the inspiralling black holes only from the early
inspiral phase when the two black holes are clearly separated and
infer the area of the final black hole only from its simple ringdown
behaviour as given by the Kerr metric.

Another result that shows up
naturally is an independent estimate of the energy radiated away
during the merger in the form of gravitational radiation without using
any of the aforementioned numerical relativity fits.  Since we obtain
the two initial masses and the final mass, it is straightforward to
compute the difference and obtain a posterior distribution of the
radiated energy.  This can be compared with the numerical relativity
prediction and thus offers yet another test of general relativity.

The various details involved in this calculation will be described in the following sections. 
The plan for the rest of the paper is as follows. 
Sec. \ref{sec:preliminary} introduces preliminary material and notation.
Sec. \ref{sec:method} presents the details of the
test. Sec. \ref{sec:example} and Sec. \ref{sec:violations} present results on simulated signals and finally
Sec. \ref{sec:conclusion} has concluding remarks on the future
prospects of this test.

\section{Preliminaries}
\label{sec:preliminary}

\subsection{Testing the assumptions}

A test of the black hole area increase theorem is a test of whether the
assumptions that go into the theorem's proof are valid. Any violation of area
increase would be proof that at least one of the assumptions does not
hold. In the test proposed here, we measure the change in area between
two asymptotic states, well before the merger and well after the
merger, where the relevant black holes can be approximated by Kerr
black holes. It is therefore not a test that the total horizon area is
increasing at all times during the coalescence. However, the change in
area between these asymptotic states can be written as a time integral
over the instantaneous rate of area change and hence our test depends
on the assumptions used to show that the rate of area change should
always be positive. A decrease in the area between asymptotic states
would be a demonstration that at least one of these assumptions was
violated (although clearly if the overall area change is positive this
does not necessarily preclude that the area was decreasing at some
point during the merger, or indeed that some of the assumptions were
mildly violated but the area still increases).

There exist in the literature several different proofs of the area
increase law that can be classified according to which type of horizon
they refer to, the main classes being event horizons and quasi-local
horizons based on the notion of marginally trapped surfaces/apparent
horizons. The proofs of the area increase law make different assumptions
in either of the two cases. In the asymptotic states considered here, both when
the two black holes are far apart and at late times when the final
black hole is in equilibrium, there is no difference between the areas
of the event horizon or quasi-local horizons.  Thus if the overall
area change is measured to be negative then this would be a violation
both of proofs using event horizons and of those using quasi-local horizons,
and therefore at least one of the common assumptions would most likely
be violated (although it is logically possible that different assumptions
are violated in the two cases).

Proofs of the area increase law for event horizons (see
e.g. \cite{HawkingEllis}) rely on three main ingredients:

i) The null curvature condition, which says
that the Ricci tensor $R_{\mu\nu}$ must satisfy
$R_{\mu\nu}\xi^\mu\xi^\nu\geq 0$ for any null vector field
$\xi^\mu$.  While the area theorem does not depend on the Einstein equations,
in Einstein's general relativity this assumption is equivalent
to the requirement that the stress-energy tensor $T_{ab}$ satisfies
the null energy condition $T_{\mu\nu}\xi^\mu\xi^\nu \geq 0 $.

ii) Asymptotic flatness and additional global conditions which ensure
that the spacetime outside the black hole (including the event
horizon) must be predictable from suitable data on a Cauchy
surface. In particular, these conditions rule out the presence of
naked singularities (cosmic censorship).

iii) The proofs use properties of the intrinsic geometry of event
horizons and in particular the geodesic deviation equation for null
geodesics on the event horizon.

It can be shown that given the previous conditions, the congruence of
null geodesics generating the event horizon cannot have negative
expansion anywhere and the area is always increasing.  The most
general statement and proof of the area increase law is by Chrusciel
et al \cite{Chrusciel:2000cu}.  This includes the cases when the event
horizon is not smooth\footnote{Generically event horizons are not
  smooth; cusps are formed when null geodesics enter the horizon
  \cite{Chrusciel:1996tw}. Numerical relativitists assume that the
  horizon is regular except for a finite number of such cusps (see
  e.g. \cite{Thornburg:2006zb}).} and also applies to non-zero values
of the cosmological constant.  As expected, cosmic censorship and the
null curvature condition are still required.

Since gravitational wave observations only probe a finite region of
spacetime, it would seem unreasonable that we could say anything
definite about naked singularities \emph{anywhere} in the universe
based on these observations.  This is also an artefact of the well known
global and teleological features of the event horizon.  Thus, should
we have observational evidence that the area increase law is violated,
we can expect the energy condition to be the main culprit.

This can be seen clearly in alternate formulations of the area
increase law.  It is possible to formulate the area increase law for
black holes in a quasi-local framework, without these global
assumptions, relying on marginally trapped surfaces and the associated
notions of dynamical and trapping horizons
\cite{Ashtekar:2004cn,Booth:2005qc,Gourgoulhon:2008pu}.  Using the
Einstein equations on the horizon, it is possible to obtain a
``physical process'' version of the area increase law which relates
the increase in area to the fluxes of in-falling matter and radiation
\cite{Ashtekar:2003hk}.  These fluxes are manifestly positive if the
dominant energy condition holds.

It is in fact known that
energy conditions can be violated in nature by a number of mechanisms
\cite{Barcelo:2002bv}. The question of whether sufficient energy
condition violation occurs during a binary black hole merge to cause
the horizon area to decrease is one of the main motivations for our
test.

In our test of the area increase law, we shall assume that the black
hole no-hair theorem holds, i.e. that any astrophysical
\emph{stationary} black hole is completely described by its mass $M$
and angular momentum $J$ and given by the Kerr solution.  This applies to the two initial black holes
and to the final black hole formed as a result of the coalescence.
Thus, any cross-section of the black holes at early and late times has
an area $A$ given by
\begin{equation}
\label{eq:area} 
A = 8\pi M^2\left(1 + \sqrt{1-\chi^2}\right)\,.
\end{equation}
Here $\chi = J / M^2$ is the dimensionless spin, which can take
values $-1 \leq \chi \leq 1$. It is an important goal
to test whether the no-hair theorems are valid and in fact, there is a
large body of work on possible tests of the black hole uniqueness
theorem from binary black hole observations~\cite{PhysRevD.85.024018, Dreyer:2003bv, Berti:2009kk, PhysRevD.85.124056, Yunes:2016jcc, baibhav2018black}.
It would be preferable
to drop this assumption and design a joint test of both the area
increase law and the uniqueness theorems, but this is beyond the scope
of the present paper.  It is important for us to clearly identify the
parts of the waveform when we cannot assume the no-hair theorem to
hold and this will be addressed below in Secs. \ref{sec:inspiral} and
\ref{sec:ringdown}.

As shown by Hawking \cite{Hawking:1971tu}, the constraint derived from the
area increase law can be used to bound the amount of energy emitted
during the coalescence process.  For a coalescence of non-spinning,
equal-mass black holes, this bound limits the emitted energy to be no
more than about 29\% of the initial rest-mass energy of the two black
holes (see also \cite{Smarr:1976qy}). Numerical simulations of black
hole collisions show that in reality the amount of energy radiated by
gravitational waves is considerably less than this theoretical upper
bound \cite{Healy:2015mla}.

\subsection{Estimating the parameters}
\label{sec:pe}

To extract the mass and spin parameters from the observations we must
examine the nature of the gravitational waves produced.  Consider a
plane gravitational wave corresponding to a gravitational metric
perturbation $h_{\mu\nu}$.  We can find a frame transverse to the
direction of propagation so that the transverse-traceless part of
$h_{\mu\nu}$ can be written in terms of two polarisations:
\begin{eqnarray}
  h_+(t) &=& A_+(t)\cos\Phi(t) \,,\\
  h_\times(t) &=& A_\times(t)\sin\Phi(t) \,.  
\end{eqnarray}
Here $A_{+,\times}$ are slowly varying amplitudes and $\Phi(t)$ is a
rapidly varying phase. The amplitudes depend on the intrinsic parameters
of the source, the distance to the binary, and the angle between
the line-of-sight vector from the binary to Earth and the orbital
angular momentum vector (inclination angle $\iota$). 

The response of an interferometric detector to
this wave depends on the relative orientation between the wave frame
and the detector frame and thus is specified by three angles.  These
three angles are typically taken to be the sky-location of the source
given by a right ascension $\alpha$ and declination $\delta$ in a
geocentric coordinate system, and the so-called polarisation angle
$\psi$ defining the relative orientation of the wave frame with the
geocentric coordinate system (see
e.g. \cite{Apostolatos:1994mx,Jaranowski:1998qm}).  Assuming the
wavelength of the signal to be much larger than the detector arms,
appropriate for ground based detectors and for the kind of signals we
are considering, the strain $h(t)$ observed by the detector is
\begin{equation}
  h(t) = F_+(\alpha,\delta,\psi)h_+(t-t_0,\phi_0) + F_\times(\alpha, \delta, \psi)h_\times(t-t_0,\phi_0)\,.  
\end{equation}
Here $F_{+,\times}$ are the beam pattern functions of the detectors
(assumed to be constant over the duration of the signal), $t_0$ is a
suitably defined arrival time and $\phi_0$ is the phase at $t_0$.  It
is useful to rewrite $h(t)$ as
\begin{equation}
  h(t) = A(t)\cos(\phi_0 + \phi(t-t_0))\,,
\end{equation}
where $A(t)$ is a slowly varying amplitude and $\phi(t-t_0)$ a rapidly
varying phase. 

We estimate the source parameters of a gravitational wave signal using
Bayesian inference, which was the method used to estimate the
parameters of GW150914 and all subsequent events
\cite{TheLIGOScientific:2016wfe}.  We wish to know the properties of
the source of a gravitational-wave signal that exists in some given
data $s(t)$. To that end, we use a model of the signal $h$ that is
parametrised by the source properties
$\{M_1, \chi_1, ...\} \equiv \gwparams$.  We cannot measure these
properties to infinite precision. Instead we seek the probability
density function $p(\gwparams|s,h)$, which quantifies our measurement
uncertainty. According to Bayes' Theorem, $p(\gwparams|s,h)$ (known as
the \emph{posterior} distribution) is proportional to the likelihood
$\mathcal{L}(s|\gwparams,h)$ of observing the data given $\gwparams$
times a \emph{prior} probability distribution $p(\gwparams)$. The
prior represents our knowledge of \gwparams{} before observing the
data.  In a network of $N_d$ gravitational-wave detectors containing
uncorrelated stationary Gaussian noise, the likelihood function is:
\begin{equation}
\label{eqn:likelihood}
\mathcal{L}(s|\gwparams,h) \propto
    \exp\left[-\frac{1}{2} \sum_{a=1}^{N_d}
              \ip{h_a(\gwparams)-s_a}{h_a(\gwparams)-s_a}
        \right],
\end{equation}
where $s_a$ is the data in each detector and $h_a$ is the model waveform (or
\emph{template}) as it would be observed in each detector. The inner product
$\ip{\cdot}{\cdot}$ is:
\begin{equation}
\label{eqn:inner_product}
\ip{x}{y} \equiv 4 \Re \int_0^\infty
    \frac{\tilde{x}_a^*(f)\tilde{y}_a(f)}{S_n^{(a)}(f)}\mathrm{d}f\,.
\end{equation}
Here $S_n^{(a)}(f)$ is the single-sided power spectral density
(PSD) of the noise in the $a$-th detector.

Stochastic samplers based on Markov-chain Monte Carlo (MCMC)
techniques can be used to evaluate Eq. \eqref{eqn:likelihood} over the
large, multidimensional space of possible parameters and produce an
estimate of $p(\gwparams|s,h)$. This estimate can then be numerically
marginalised to yield Bayesian credible intervals on various
parameters.

\section{Method}
\label{sec:method}

Normally, when analysing the entire signal (and assuming the detector noise truly
is stationary and Gaussian), the parameter estimates produced by the method
described in Sec. \ref{sec:pe} will be unbiased. In other words, if the source
distribution is the same as the prior, then we can expect that a signal's true
parameters will lie within the $X\%$ credible interval $X\%$ of the time.

However, the method described in Sec. \ref{sec:pe} will yield biased estimates
if a parameter-dependent cut is applied to the template waveforms
\cite{Mandel:2014tca},  as we wish to do here. This is due to the fact that Eq.
\eqref{eqn:likelihood} is derived assuming that the template is a model for the
entire signal; i.e., $\mathcal{L}(s|\gwparams,h)$ is maximised for parameters
that best match the signal averaged over the entire bandwidth of the detector.
Here, however, we wish to find the parameters that best-match the signal only
over a limited portion of the signal, while ignoring the rest. Crucially, the
onset and duration of the part that we wish to exclude --- the merger --- is
dependent on the signal's source parameters. Since this additional,
parameter-dependent condition is not included in $p(\gwparams|s,h)$, a na\"ive
application of Eq. \eqref{eqn:likelihood} results in biased measurements, as
we will see below.

Further complicating our efforts is our desire to excise a part of the signal
in the time domain rather than the frequency domain. As discussed below, this
effectively couples the sky location of the signal to the measurement of the
initial and final masses and spins.

In the following we describe a method to overcome these challenges. To
illustrate and test the method, we simulate a signal using a publicly available
waveform produced by the SXS collaboration~\cite{SXS:catalog, Mroue:2013xna}.
We choose a non-spinning, equal-mass binary black hole\footnote{Numerical
waveform SXS:BBH:0066, Lev5} with total mass $M_t=70 M_\odot$.\footnote{Since
we use waveform templates that do not include sub-dominant modes, we select
only the 22 mode of the numerical waveform. This is a sensible approach because
higher modes are not expected to be detectable with current 
sensitivities~\cite{PhysRevD.85.024018, PhysRevD.90.124032}.}
The signal, which is similar to a signal like GW150914, 
is injected in zero noise at a luminosity distance $D_L = 500\,$Mpc.

We compare results using two different detector sensitivities: the PSD
published in the LIGO Open Science Centre~\cite{LOSC} for GW150914, which is
representative of Advanced LIGO's first observing run, and the zero-detuned
high-power (ZDHP) PSD \cite{ZDHP}, which is representative of LIGO's expected sensitivity
in the coming years. In this analysis we only consider the two LIGO detectors.
However, the method can be trivially expanded to include any number of
detectors. 

We use the PyCBC Inference framework to estimate $p(\gwparams|s,h)$
\cite{pycbcinf}.  This is a python-based pipeline similar to the LALInference
pipeline \cite{Veitch:2014wba} used to infer the parameters of published
gravitational waves. PyCBC Inference supports multiple stochastic sampling
engines. In this study we use \kombine{} \cite{kombine}, which is an MCMC
sampler that uses an ensemble of Markov chains (or \emph{walkers}) to
efficiently estimate $p(\gwparams|s,h)$.

\subsection{The effect of sky location}

An integral part of our analysis is separating
the inspiral part to estimate the initial parameters of the binary and
the ringdown part to estimate the parameters of the final black
hole. The templates used for the parameter estimation are terminated (or
started) at a specific time. This method complicates the issue of
dealing with the sky location since the sky location affects the
arrival time of the signal in the detectors.

We have found that if the sky location is allowed to vary in the MCMC, the
terminated templates will favour sky locations that get them closer to the
merger, yielding biased results. 
One could fix the sky location to a single point if it were known.  However, we
do not expect to measure the sky location of a binary black hole merger to
sufficient precision with the current network of gravitational-wave detectors
for this approach to work. Indeed, when doing the ringdown analysis, we have
found that choosing different fixed points drawn from within the $50\%$
credible interval of the sky location produced by a full
inspiral-merger-ringdown analysis yields statistically significant different
estimates of the final mass and spin. Thus picking a single point for the sky
location when one is not actually known will result in an underestimate of our
uncertainty, and a potential bias, of these parameters.

To account for this uncertainty, we fix the sky location to a distribution
rather than a single point when doing the inspiral and ringdown analyses. We do
this by assigning each walker in our MCMC to a different right ascension and
declination. These locations are drawn from a given sky map that quantifies the
uncertainty in the event's location. The walker's positions in the sky remain
fixed throughout the entirety of the parameter estimation routine. This way, we
include all the information obtained from allowing the sky location to vary
without forcing the entire parameter estimation analysis to remain on one fixed
point.

We use the same sky locations for the inspiral and ringdown analysis. After
evolving the MCMC until it is burned in, each walker in the inspiral (ringdown)
analysis produces a point estimate of the initial (final) area. Taking the
estimates over all of the walkers produces a distribution of areas. When taking
the difference in areas, we only compare point estimates between the same sky
locations in both runs. Thus, the inspiral and ringdown analyses are
independent in all parameters except for the sky location.

\begin{figure}[tb]
\centering
\includegraphics[width=\columnwidth]{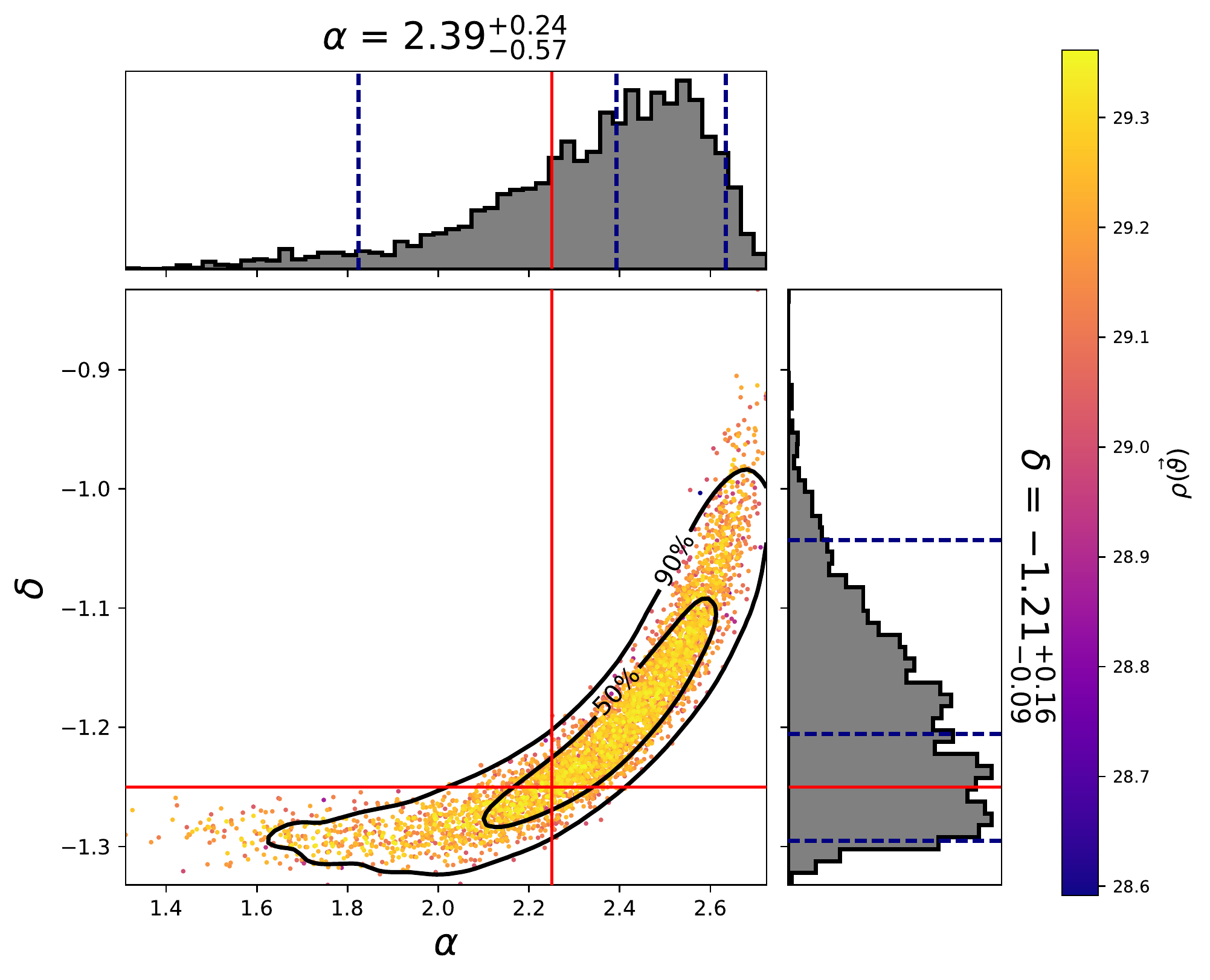}
\caption{\label{SkyLoc} Posterior distribution for the sky location (right ascension $\alpha$ and declination $\delta$) obtained from the full simulated signal. The colorbar shows the signal-to-noise ratio (SNR), which is a function of the likelihood, and the red lines indicate the values that were injected. The centre and outer dashed lines in the histograms represent the median value and the 90\% credible interval, respectively. These correspond to the values and the errors given on top of the histograms.}
\end{figure}

To produce a sky map for this study, we first perform the parameter estimation
analysis on the full signal, using full inspiral-merger-ringdown IMRPhenomD
templates~\cite{PhysRevD.93.044006, Khan:2015jqa}. Figure~\ref{SkyLoc} shows the marginalised
posterior distribution of the right ascension ($\alpha$) and declination
($\delta$) obtained from this analysis. We only use the sky-location
information from this analysis; no information regarding the GW parameters
$\gwparams$ are kept.
Here, we have restricted our analysis to the two
LIGO detectors. However it is trivial to include a third detector, if
available, to improve our measurement of the sky
location~\cite{GW170814, GW170817}.

Since the sky location relies primarily on the time of arrival of the signal at
different detectors, it is possible to estimate the sky location using searches
that do not use any waveform models from general relativity. For example, a sky
map produced using the coherent Wave Burst pipeline \cite{Klimenko:2015ypf} was
published with GW150914 \cite{LOSC}. The sky map produced by this pipeline may
be used when applying our test to real events.

\subsection{The inspiral analysis} 
\label{sec:inspiral} 

The initial parameters ought to be measured from the inspiral part of the
waveform, with no assumptions or input from the merger or ringdown.  Ideally,
one would use post-Newtonian waveforms~\cite{Buonanno:2009zt}. However, full
inspiral-merger-ringdown waveforms such as the IMRPhenomD
family~\cite{PhysRevD.93.044006, Khan:2015jqa} have better agreement with numerical relativity
waveforms at higher inspiral frequencies. 
We shall use IMRPhenomD waveforms terminated at the end of the inspiral 
to exclude the highly dynamical merger phase and guarantee that the 
inspiral analysis is completely independent from the merger and ringdown. 
For compact binaries with arbitrary masses and spins, the 
hybrid minimum energy circular orbit (hybrid MECO)~\cite{Cabero:2016ayq} is a 
proxy for the end of the inspiral. The hybrid MECO depends on the mass ratio of
the binary and on the spins of the black holes, and is symmetric under exchange
of the individual objects' parameters.

Na\"ively, one might think that to exclude the merger phase from the analysis,
we need only to terminate the templates at the time at which they pass through
hybrid MECO. However, while this excludes the merger dynamics in the templates,
it does not exclude the merger in the signal. This results in templates with
higher spin and mass matching the signal better than the template that has the
same mass and spin of the signal, leading to biased results. This is because
the hybrid MECO of these templates occurs at a higher frequency. Effectively,
these templates are able to see more of the signal; the additional
signal-to-noise ratio (SNR) they gain in doing so is enough to offset any
mismatch these templates have with the signal at lower frequencies.

To recover the correct masses and spins it is necessary to exclude the merger
dynamics in both the templates and the signal. However, because the intrinsic
parameters of the signal are unknown, the time at which the signal passes
through hybrid MECO is also unknown.  To estimate this time we choose a grid of
times $\tgrid < \tref$, where \tref{} is a fiducial time chosen arbitrarily in
the proximity of the expected coalescence time of the binary. We perform an
independent parameter estimation analysis for each grid time. In each analysis,
we apply a taper function to the templates in the time domain that goes to zero
at a time \ttaper{}. The taper time is varied across parameter space and
between detectors. Specifically, for a given set of parameters \gwparams{} and
a detector $\mathrm{D}$,
\begin{equation}
\label{eqn:ttaper}
\ttaper = \min\left[\tgrid + \delta t(\alpha, \delta; \mathrm{D}),~\thmeco(\gwparams)\right],
\end{equation}
where \thmeco{} is the time at which the template goes through hybrid MECO. The
$\delta t(\alpha, \delta; \mathrm{D})$ is an offset applied to account for the
arrival time uncertainty in each detector arising from the uncertainty in sky
location.

When doing the analysis we found that whitening the template before tapering
yielded better results than simply applying the taper to the waveform and then
whitening. That is, we replace the full IMR template $h$ with:
\begin{equation*}
h'(t) = w(t; \ttaper) [h * \widetilde{A^{-1}}](t),
\end{equation*}
where $\widetilde{A^{-1}}(t)$ is the inverse Fourier transform of
$1/\sqrt{S_n(f)}$; the $*$ indicates convolution. We use half a Kaiser window
with a duration of $10\,$ms and shape parameter $\beta=8$~\cite{kuo1966system} for
the taper function $w(t; \ttaper)$.  A duration of $10\,$ms is used
because the whitening filter $\widetilde{A^{-1}}(t)$ effectively goes to zero on
this time scale, ensuring that times $t > \ttaper$ are minimally coupled to
times $t < \ttaper$ via the convolution. The whitened, tapered template $h'(t)$
is filtered with the whitened data.

The grid time \tgrid{} prevents the templates from matching the merger dynamics in the signal. For \tgrid{} times later than \thmeco{}, results are
influenced by the dynamic merger phase and the posterior distributions yield
biased results. As \tgrid{} approaches \thmeco{} of the signal, the posterior
settles around the parameters of the signal. For \tgrid{} times earlier than
the signal's \thmeco{}, the posterior remains in the same region of parameter
space, though it begins to widen due to the decreasing SNR.  Therefore, the
transition time between the moving and the growing posterior distributions
yields the best point at which to calculate the initial areas. 

It is thus clear that to estimate the transition time, we need to
study how the posterior distributions change as \tgrid{} is
varied.  A general notion of the divergence between two probability
distributions $p_1(x)$ and $p_2(x)$ is provided by the
Kullback-Leibler divergence (see e.g. \cite{CoverThomas1991}) which,
for discrete distributions is defined as
\begin{equation}
D_{KL} (p_1 || p_2) =  \displaystyle \sum_x p_1 (x) \log \frac{p_1 (x)}{p_2 (x)} \, .
\end{equation}
However $D_{KL}(p_1||p_2)$ is not symmetric, i.e.
$D_{KL}(p_1||p_2) \neq D_{KL}(p_2||p_1)$.  In particular, $D_{KL}$
cannot be viewed as a distance between probability distributions.
$D_{KL}$ is an appropriate divergence to use when one of the
distributions is privileged for some reason.

In our case, we have no reason to distinguish different values
of $\tref-\tgrid$ and thus we choose to use instead the Jensen-Shannon (JS)
divergence.  Given two probability distributions $p_1$ and $p_2$, the
JS divergence is a symmetric and smooth version of the
Kullback-Leibler (KL) divergence $D_{KL} (p_1 || p_2)$:
\begin{equation}
D_{JS} (p_1 || p_2) = \frac{1}{2} D_{KL} (p_1 || q) + \frac{1}{2} D_{KL} (p_2 || q) \, ,
\end{equation}
where $q = \frac{1}{2} (p_1 + p_2)$.  It has been shown that
$D_{JS}$ can be turned into a distance measure between probability
distributions \cite{EndresShindelin}.

\begin{figure}[tb]
\centering
\includegraphics[width=\columnwidth]{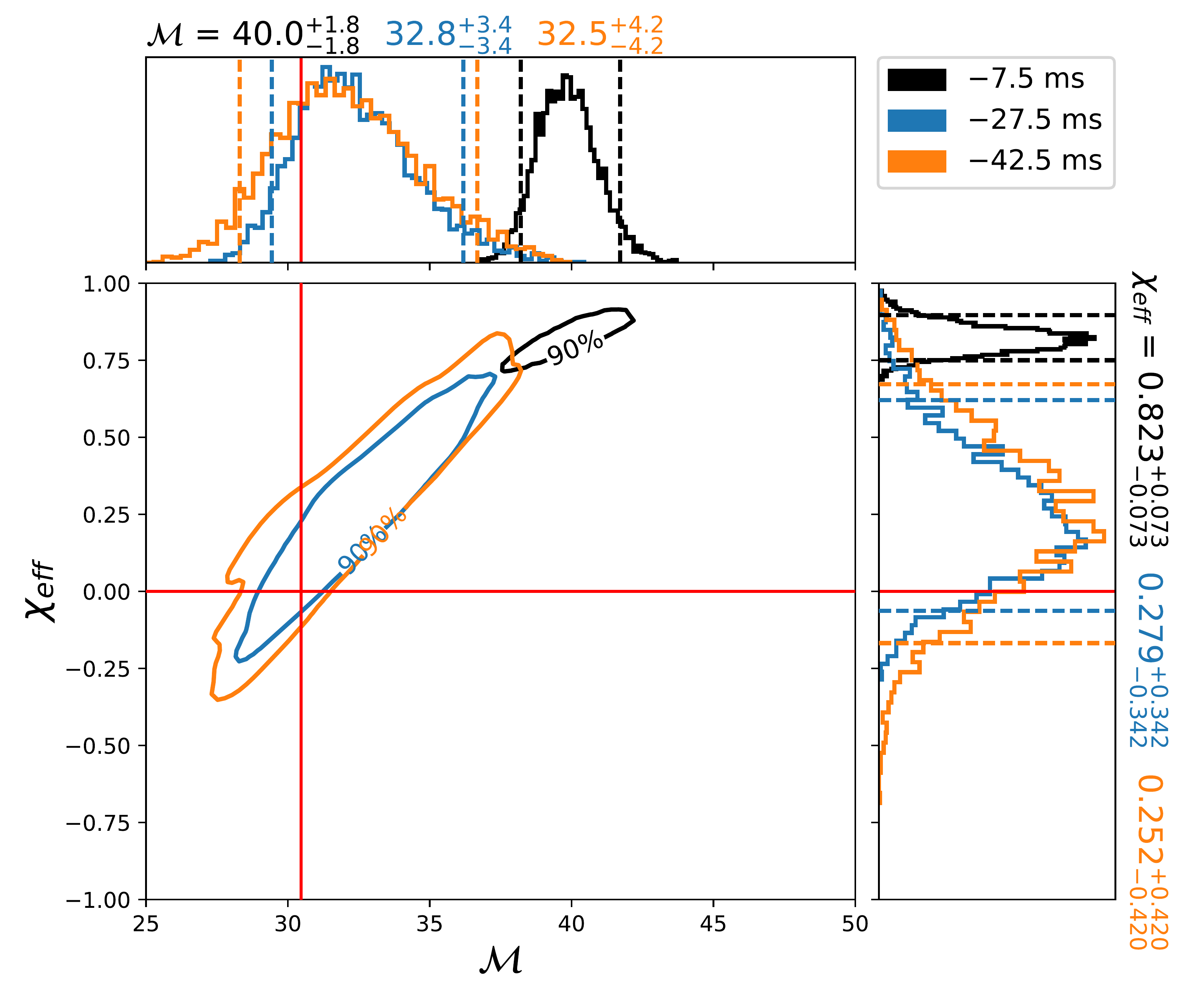}
\includegraphics[width=\columnwidth]{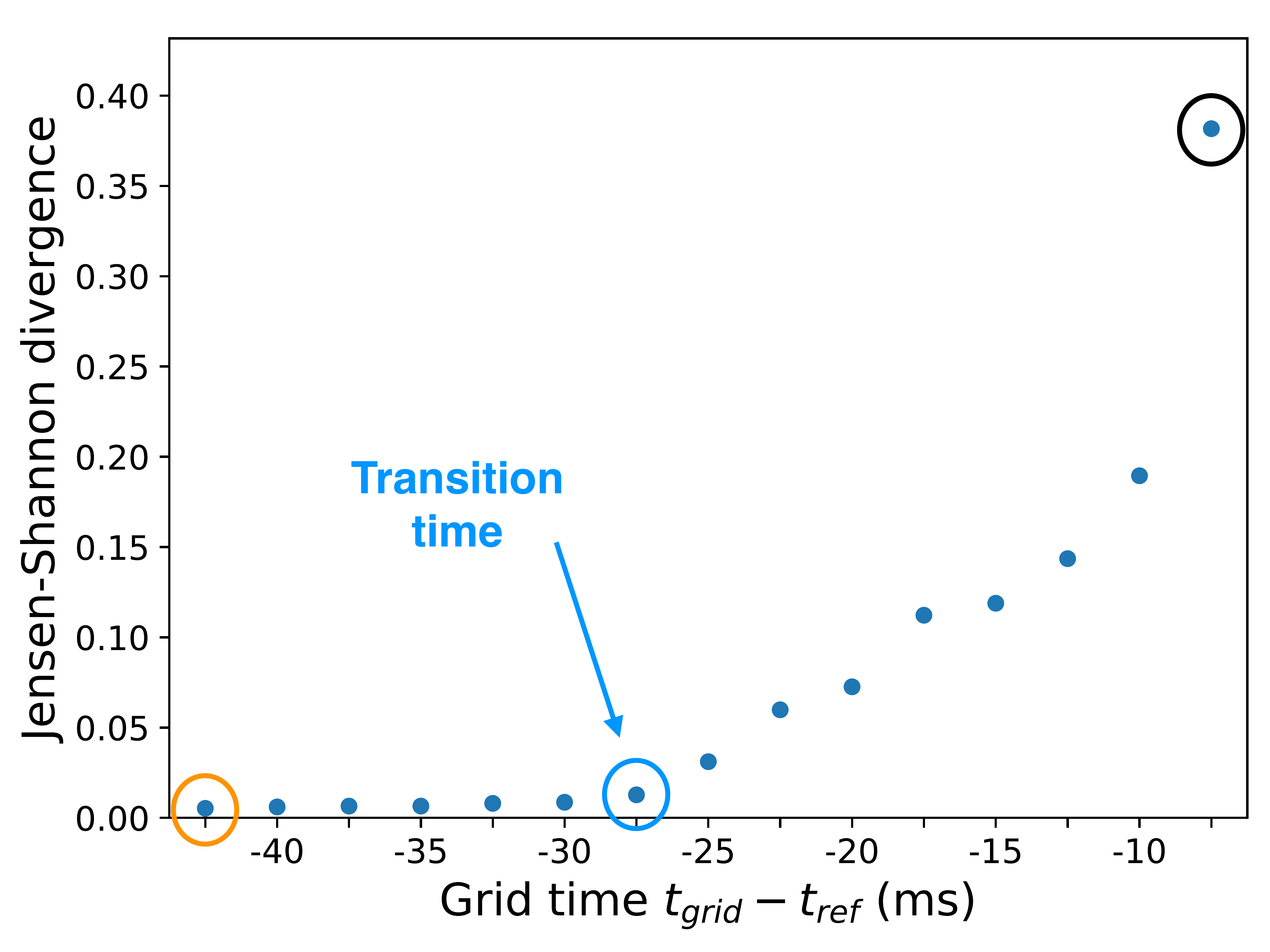}
\caption{\label{ChooseTime} (\textit{Top}) Posterior distribution of the
  chirp mass and the effective spin obtained from three different runs.
  The red cross indicates the injected values.
  (\textit{Bottom}) Jensen-Shannon divergence between
  consecutive grid runs. The $x$-axis
  indicates the corresponding time $\tgrid$ and the JS
  divergence is calculated between the posterior distributions at 
  $\tgrid$ and $\tgrid + \Delta t$, where 
  $\Delta t = 2.5$ ms. The coloured circles indicate the times
  corresponding to the posterior distributions in the top figure.
  The chosen transition time is $\tgrid - \tref=-27.5$ ms.}
\end{figure}

Figure~\ref{ChooseTime} shows the JS divergence between the posterior
distribution of the analysis at \tgrid{} and the analysis at $\tgrid  + \Delta t$, 
which we have chosen to be $\Delta t = 2.5$ ms. Since we are
only interested in the masses and spins of the black holes, we compute the JS
divergence using the 2D marginalised distribution of chirp mass $\mathcal{M} =
(M_1 M_2)^{3/5}/M_t^{1/5}$ and effective spin $\chi_{\text{eff}} = (M_1 \chi_1
+ M_2 \chi_2) / M_t$. While the posterior distribution is still moving, the JS
divergence is changing significantly. The transition time is given by the maximum
inspiral (minimum ringdown) grid
time where the JS divergence is still roughly constant. From
Figure~\ref{ChooseTime}, the transition time is clearly seen to be at the time
$\tgrid = \tref - 27.5$ ms. At times closer to the merger, the
difference between consecutive JS divergences increases significantly.

In doing the analysis shown in Fig.~\ref{ChooseTime}, we vary the two component
masses $M_{1,2}$, two component spins $\chi_{1,2}$, distance $D_L$, inclination
$\iota$, polarisation $\psi$, coalescence phase $\phi_0$, and the template's
coalescence time $t_c$. We use uniform priors for all of these parameters. The
prior range for the parameters of interest is $M_i \in [10, 80)\,M_{\odot}$ and
$\chi_i \in [-0.9895, 0.9895)$. This is the same prior that was used in the
analysis of GW150914 \cite{TheLIGOScientific:2016wfe}.

\subsection{The ringdown analysis} 
\label{sec:ringdown} 

The late ringdown phase is well described through perturbation theory. Assuming
that the final object is a Kerr black hole, the ringdown signal consists of a
sum of exponentially damped sinusoids~\cite{0264-9381-26-16-163001}. 
It is natural to write the gravitational
wave in terms of spin-weighted spheroidal harmonics: 
\begin{equation} 
h_+ + ih_\times = \sum_{\ell,m,n} {}_{-2}S_{\ell m}(\iota,\varphi) A_{\ell mn}
e^{i (\Omega_{\ell mn}t + \phi_{\ell m n})}\,.  
\end{equation} 
The sum in the
above equation is over the quantum numbers $\ell,m$ and the overtone $n$. Thus,
$m = -\ell, -\ell + 1, \ldots, 0 , \ldots, \ell$ for each $\ell = 2, 3,\ldots$
and $n=0,1,2,\ldots$.  The angular functions ${}_{-2}S_{\ell m}(\iota,\varphi)$
are the spin-weighted spheroidal harmonics which appear as the angular
eigenfunctions of the equations describing the gravitational perturbations of a
Kerr black hole found by Teukolsky (see e.g. \cite{Chandrasekhar:1985kt}).
These functions depend on the angular momentum and mass of the Kerr black hole
and reduce to the usual spin-weighted spherical harmonics for the non-spinning
case. The inclination angle $\iota$ is the angle between the line-of-sight
vector from the black hole to Earth and the intrinsic angular momentum of the
black hole, and $\varphi$ is the azimuth angle of the black hole with respect
to the observer.  The amplitudes $A_{\ell mn}$ and the phases $\phi_{\ell m n}$
are arbitrary as far as our ringdown analysis is concerned.
In principle they depend on the configuration of the gravitational perturbation
$h_{\mu\nu}$ at the beginning of the ringdown phase (when linear perturbation
theory begins to be applicable) which in turn depends on the initial
configuration of the binary and on the particular gravitational theory.
However, we shall make no assumption relating the amplitudes or phases to the
initial parameters of the binary.  The complex frequencies $\Omega_{\ell m n}$
are the quasi-normal frequencies determined from the Teukolsky equation. They
are the frequencies for which we obtain solutions which are purely outgoing at
infinity and purely ingoing at the horizon.  See \cite{Leaver:1985ax} for a
method of calculating these frequencies for a Kerr black hole and see e.g.
\cite{Kokkotas:1999bd,Berti:2009kk} for reviews on black hole ringdown.

The two polarisations of the gravitational waveform are given by \begin{align}
\label{td_ringdown} h_+(t) &= \sum_{\ell,m,n} {}_{-2}Y_{\ell m}^+(\iota)
A_{\ell m n} e^{-t/\tau_{\ell m n}} \cos(\omega_{\ell m n} t + \beta_{\ell m
n}) \, , \nonumber \\ h_\times(t) &= \sum_{\ell,m,n} {}_{-2}Y_{\ell
m}^\times(\iota) A_{\ell m n} e^{-t/\tau_{\ell m n}} \sin(\omega_{\ell m n} t +
\beta_{\ell m n}) \, .  \end{align} It is assumed here that the ringdown begins
at $t=0$ and instead of the complex frequency $\Omega_{\ell m n}$ we have used
the damping time $\tau_{\ell m n}$ and real frequency $\omega_{\ell m n}$.
Here, we have approximated spin-weighted spheroidal harmonics ${}_{-2} S_{\ell
m n}$ by spin-weighted spherical harmonics ${}_{-2} Y_{\ell m
n}$~\cite{PhysRevD.73.024013, PhysRevD.76.104044}.  The angle $\beta_{\ell m n}
= \phi_{\ell m n} + m \varphi$ combines the initial ringdown phase with the
azimuthal part of the spherical harmonics, and we have
defined~\cite{PhysRevD.76.104044} 
\begin{align} 
{}_{-2}Y_{\ell m}^+(\iota) &= {}_{-2}Y_{\ell m}(\iota, 0) + (-1)^\ell {}_{-2}Y_{\ell \text{-}m}(\iota, 0) \, , \nonumber \\ 
{}_{-2}Y_{\ell m}^\times(\iota) &= {}_{-2}Y_{\ell m}(\iota, 0) -(-1)^\ell {}_{-2}Y_{\ell \text{-}m}(\iota, 0) \, .  
\end{align}
Throughout this paper we use only the fundamental mode ($\ell=m=2, n=0$).
However, the methods presented here can be extended to multi-mode ringdowns
once the detectors' sensitivities allow the detectability
of higher order modes.

\begin{figure}[tb]
\centering
\includegraphics[width=\columnwidth]{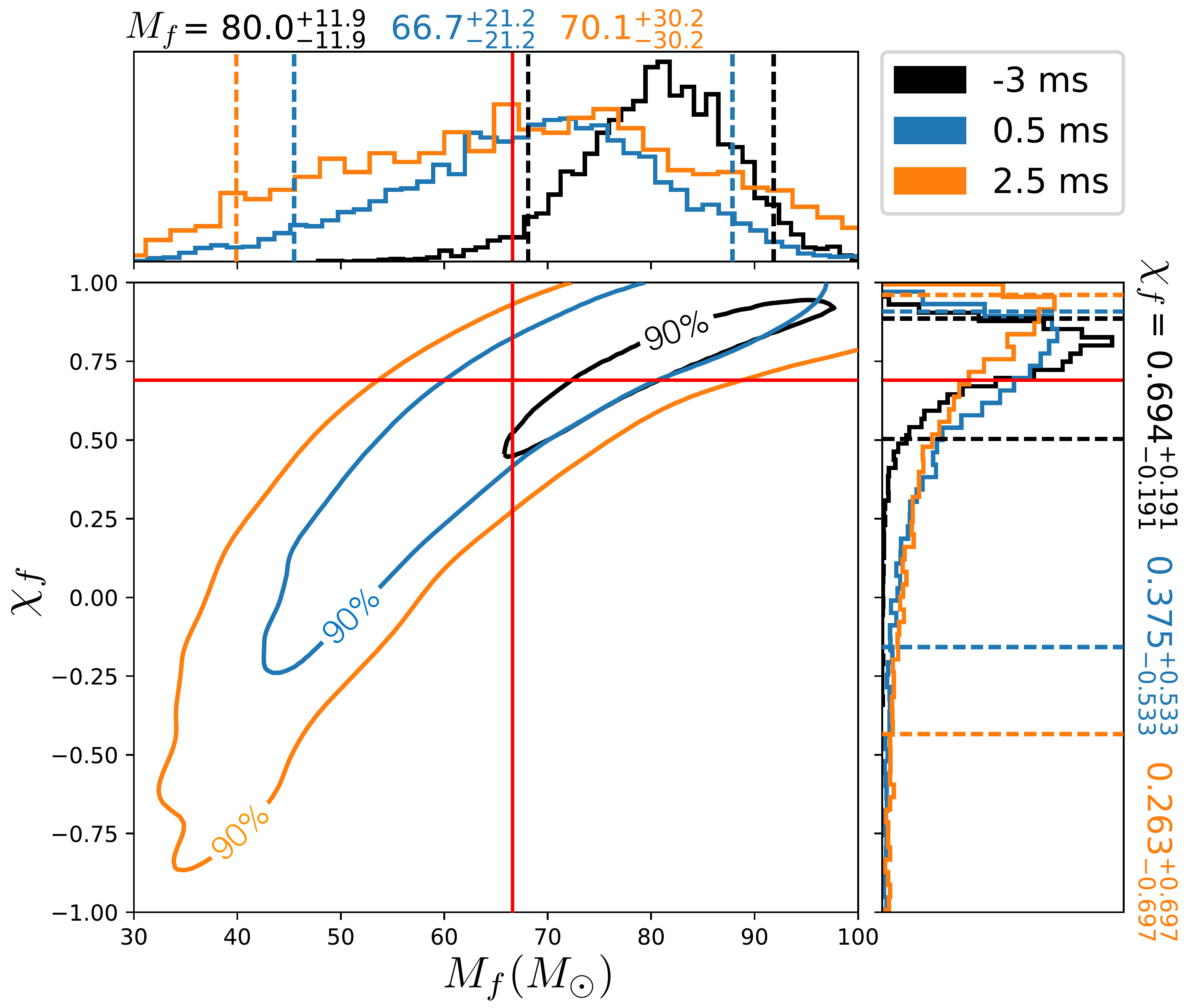}
\includegraphics[width=\columnwidth]{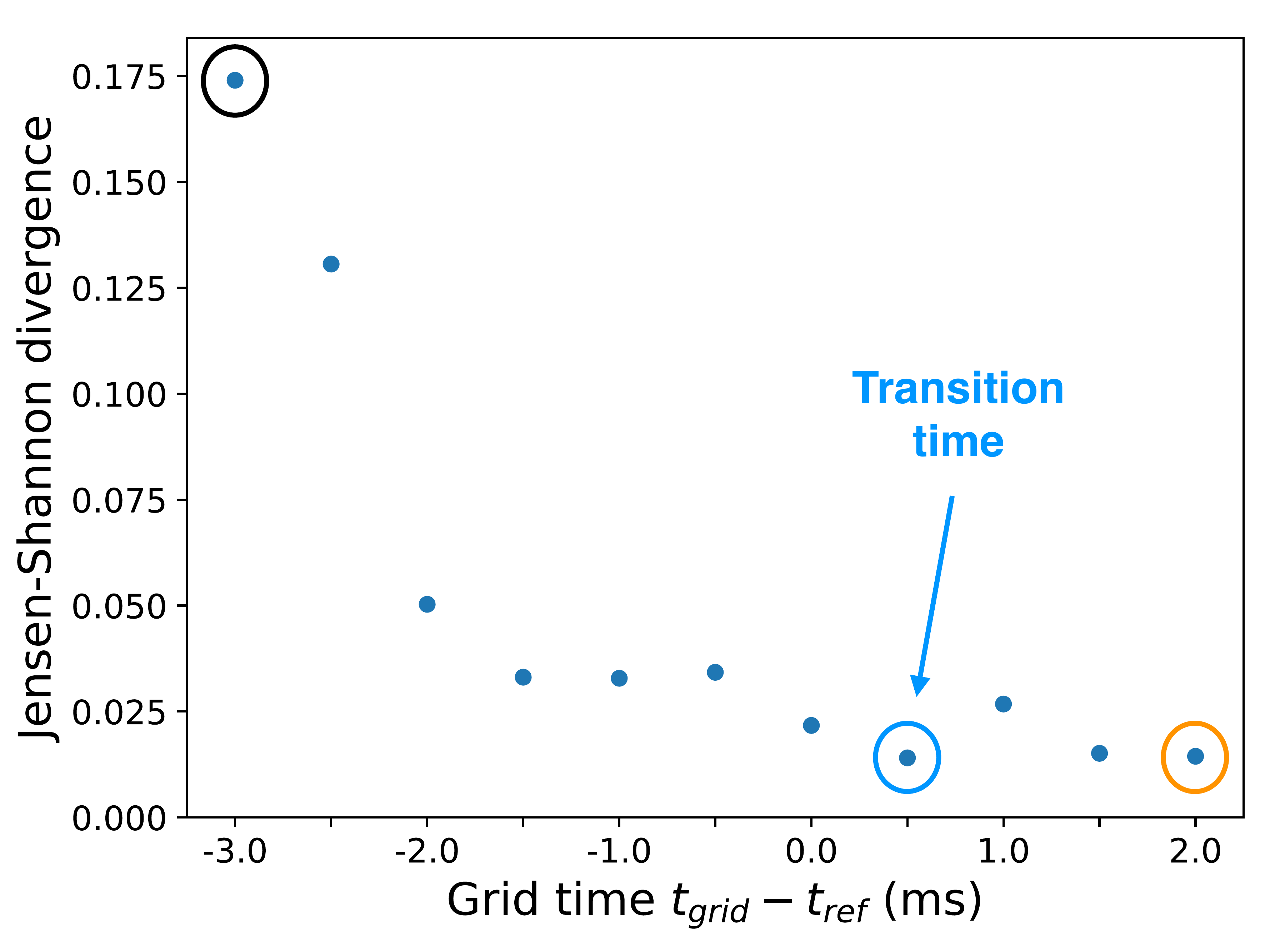}
\caption{\label{RingdownTime} (\textit{Top}) Posterior distribution of the
  final mass $M_f$ and the final spin $\chi_f$ obtained from three different runs.
  The red cross indicates the injected values.
  (\textit{Bottom}) Jensen-Shannon divergence between
  consecutive grid runs. The $x$-axis
  indicates the corresponding time $\tgrid$ and the JS
  divergence is calculated between the posterior distributions at 
  $\tgrid$ and $\tgrid + \Delta t$, where 
  $\Delta t = 0.5$ ms. The coloured circles indicate the times
  corresponding to the posterior distributions in the top figure.
  The chosen transition time is $\tgrid - \tref=0.5$ ms.}
\end{figure}

The ringdown template assumes $h(t) = 0$ before the start of the damped
sinusoid, but it is unknown in the data when the signal waveform starts
behaving like a pure damped sinusoid. Same as for the inspiral phase, we choose
a grid of times $\tgrid > \tref$ and perform the ringdown analysis for each time
separately, where \tref{} is the same fiducial time used for the inspiral
analysis. Times closer to the coalescence of the signal will again show biased
results. If \tref{} is after the coalescence time of the signal, we might have
to use a few times $\tgrid<\tref$ to find the transition time.
Figure~\ref{RingdownTime} shows the resulting JS divergence between the 
posterior distribution of the analysis at $\tgrid$ and the analysis at 
$\tgrid + \Delta t$, which for the ringdown analysis we have chosen to be 
$\Delta t = 0.5\,$ ms. The JS divergence is computed using the 2D
marginalised distribution of final mass $M_f$ and final spin $\chi_f$.
The transition time in the ringdown analysis happens at the time
$\tgrid = \tref + 0.5$ ms.

We find that it is necessary to zero out the data prior to the grid time in
order to accurately recover the final mass and spin using the damped sinusoid.
This is different than the inspiral analysis, in which only the templates were
modified. As discussed above, the time domain representation of the whitening
filter $\widetilde{A^{-1}}(t)$ has non-zero support on time scales
$\mathcal{O}(\mathrm{ms})$.  This is significant in the ringdown analysis, in which
differences of a few milliseconds can have large effects on the estimated
parameters. Namely, the convolution of the whitening filter with the signal in
Eq.~\eqref{eqn:inner_product} couples information from the merger with the
post-merger ringdown. Since the damped sinusoid is simply zero prior to its
onset, the whitening filter has a different effect on it. Thus, even if the
template and the post-merger signal are exactly the same prior to whitening,
they are different afterward. This difference particularly biases the recovered
damping time, which in turn affects the estimated final mass and spin. Zeroing
out the data prior to whitening decouples the whitened signal's merger and
ringdown, and causes the whitening filter to affect the signal and template in
the same way, correcting the bias.

The variable parameters for the ringdown analysis are the
central frequency $f_{220} = \omega_{220} / 2 \pi$, damping time $\tau_{220}$, amplitude $A_{220}$,
phase $\beta_{220}$, inclination $\iota$ and polarisation $\psi$ (we drop the
$220$ label from now onwards). We use uniform priors for all of these
parameters. Using the fitting formulae in~\cite{PhysRevD.73.064030}, one
can obtain the final black-hole's mass $M_f$ and spin $\chi_f$ from the
ringdown frequency and damping time. The prior range for the parameters of
interest is $f \in [20, 1024)$ Hz and $\tau \in [0.1, 100)$ ms, with the
further constraint that $f$ and $\tau$ have to yield physical masses and spins
(i.e. $M_f > 0$ and $-1 < \chi_f < 1$). 

\section{Combined results}
\label{sec:example} 

\begin{figure}[tb]
\centering
\includegraphics[width=\columnwidth]{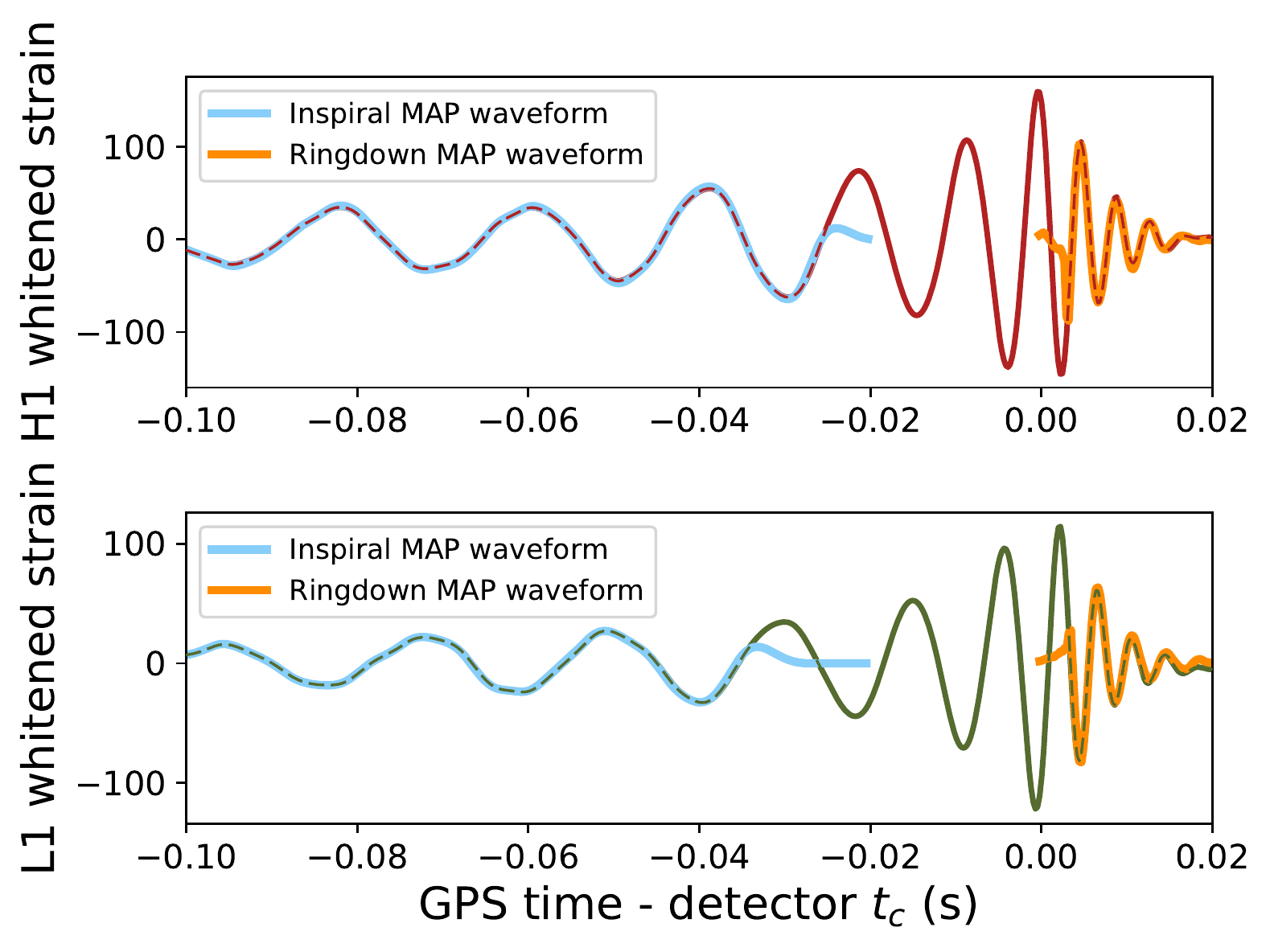}
\caption{\label{Strain} Whitened strain in each detector with the maximum posterior (MAP) waveform from the inspiral analysis (blue) and from the ringdown analysis (orange).}
\end{figure}

A visual representation of the maximum posterior (MAP) waveforms resulting from
the separated parameter estimation analyses is shown in Figure~\ref{Strain},
with the template waveforms plotted on top of the detectors' whitened strains. 
The probability distribution for the change in the area is obtained
by combining the posterior distributions of the initial and final
parameters from the selected inspiral and ringdown results.

The simulated binary black hole signal used in this paper
is bound to agree with the area theorem by design.
Using the fitting formulae in~\cite{PhysRevD.93.044006} we can estimate
the expected area increase for the signal. Two non-spinning 
black holes with masses $M_1 = M_2 = 35 M_\odot$ yield a 
final black hole with mass $M_f \simeq 66.6 M_\odot$ and spin 
$\chi_f \simeq 0.69$, which translates into an expected area increase
$A_f / A_i \simeq 1.56$.

\begin{figure}[tb]
\centering
\includegraphics[width=\columnwidth]{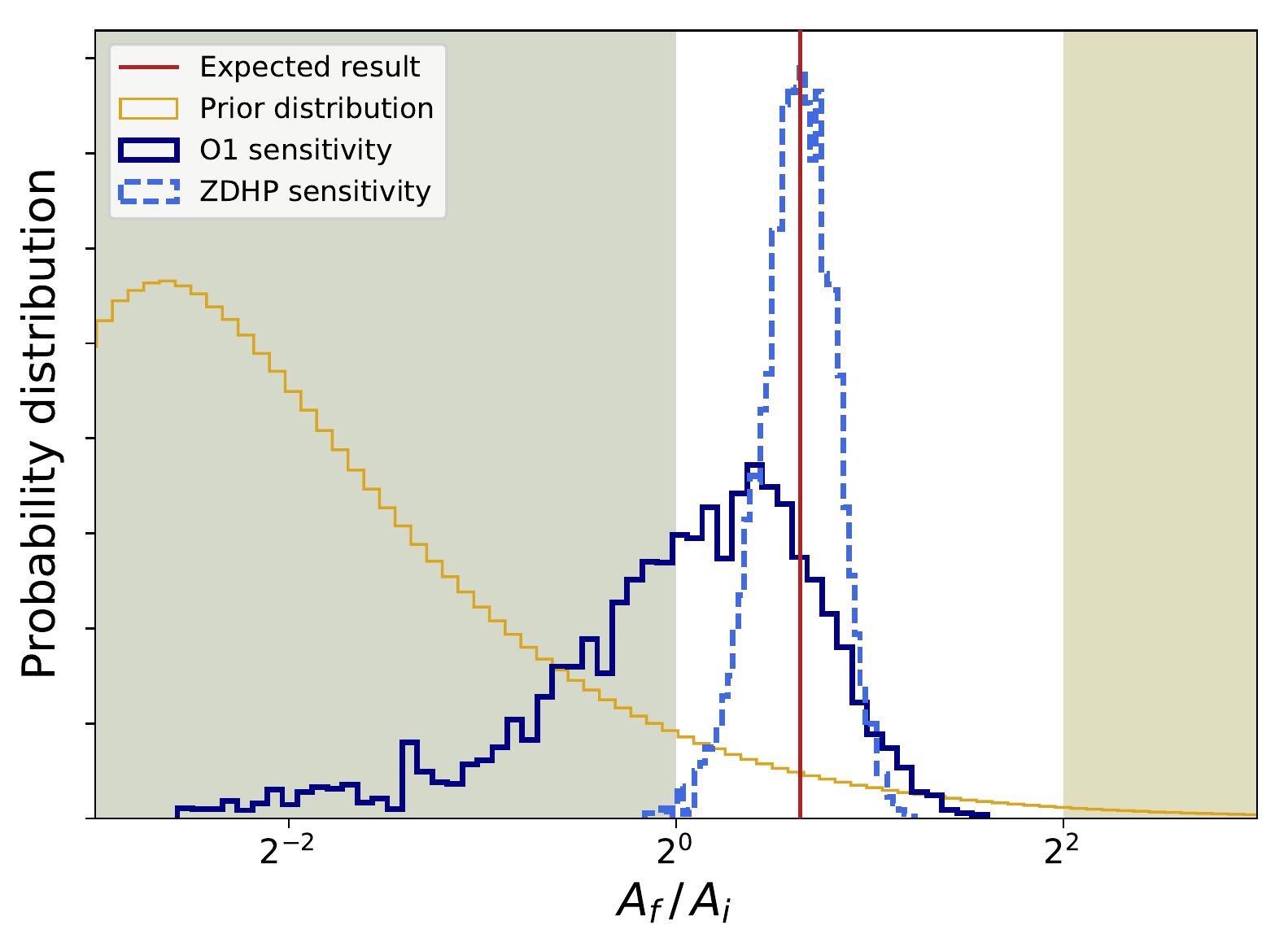}
\includegraphics[width=\columnwidth]{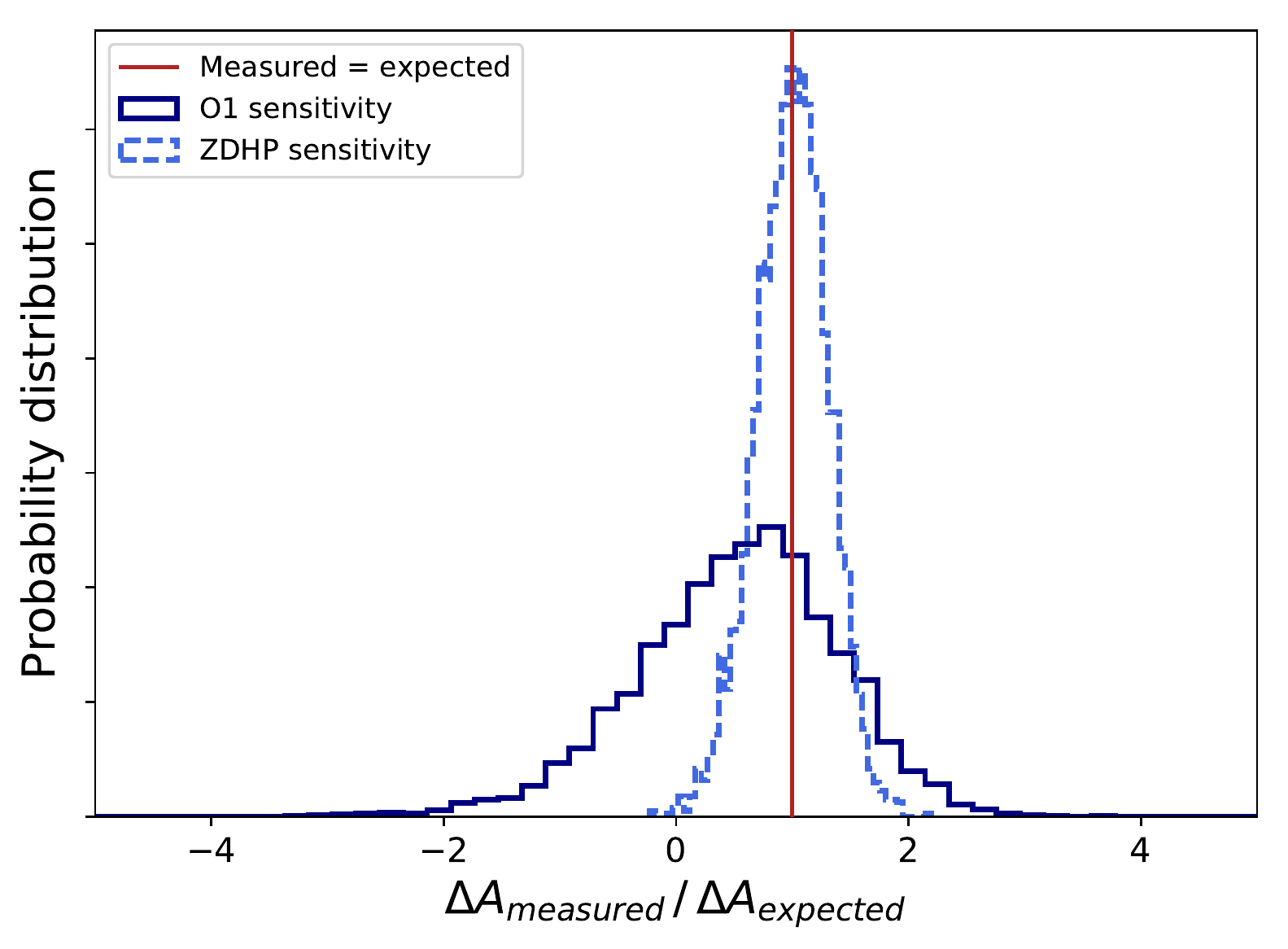}
\caption{\label{AreaResult} (\textit{Top}) Posterior distribution on the ratio
of the final to initial areas, $A_f / (A_1 + A_2)$, for two different Advanced
LIGO sensitivities, O1 and ZDHP. The shaded region $A_f/A_i < 1$ indicates
violation of the area theorem, and $A_f / A_i > 4$ indicates violation of the
conservation of energy. The vertical red line is the expected area increase.
(\textit{Bottom}) Distribution of the ratio $\Delta
A_{\text{measured}} / \Delta A_{\text{expected}}$. The measured area change
corresponds to the distribution shown in the top figure. The expected area
change is given by the initial parameters obtained in the inspiral analysis
and the corresponding expected final parameters from the fitting formulae
in~\cite{PhysRevD.93.044006}. The vertical red line indicates agreement
between the measured and the expected values, i.e. $\Delta A_{\text{measured}}
/ \Delta A_{\text{expected}} = 1$.}
\end{figure}

The top panel in Figure~\ref{AreaResult} shows the 
measured area increase, with the expected
value indicated by a red line. 
With current gravitational-wave detectors sensitivities (O1), the measured 
median value with 90\% credible
interval is $A_f / A_i = 1.31_{-0.70}^{+0.84}$. Furthermore, we obtain 
a $\sim74.6\%$ probability that the simulated signal is consistent 
with the area theorem.

\begin{figure}[tb]
\centering
\includegraphics[width=\columnwidth]{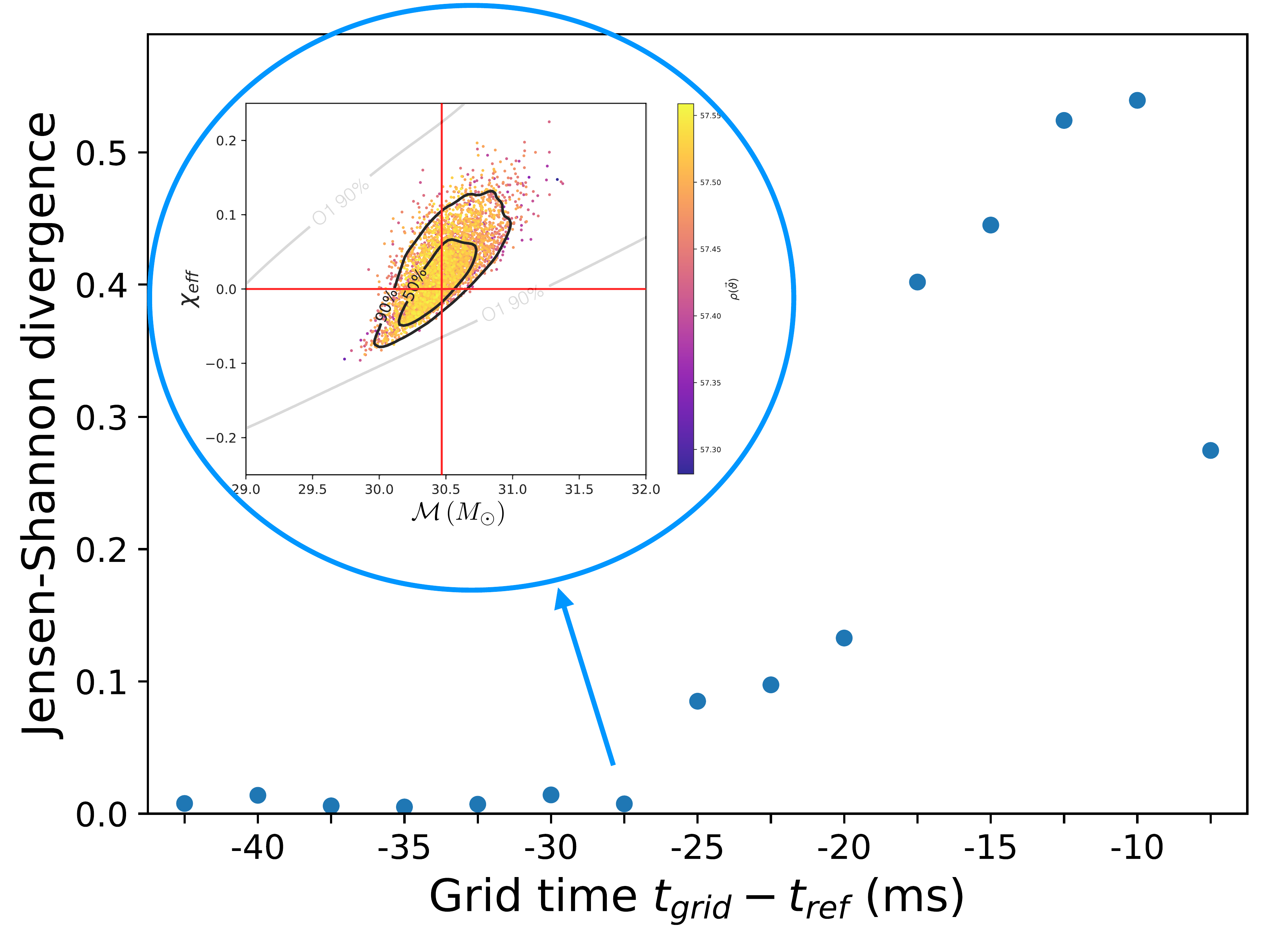} \\
\includegraphics[width=\columnwidth]{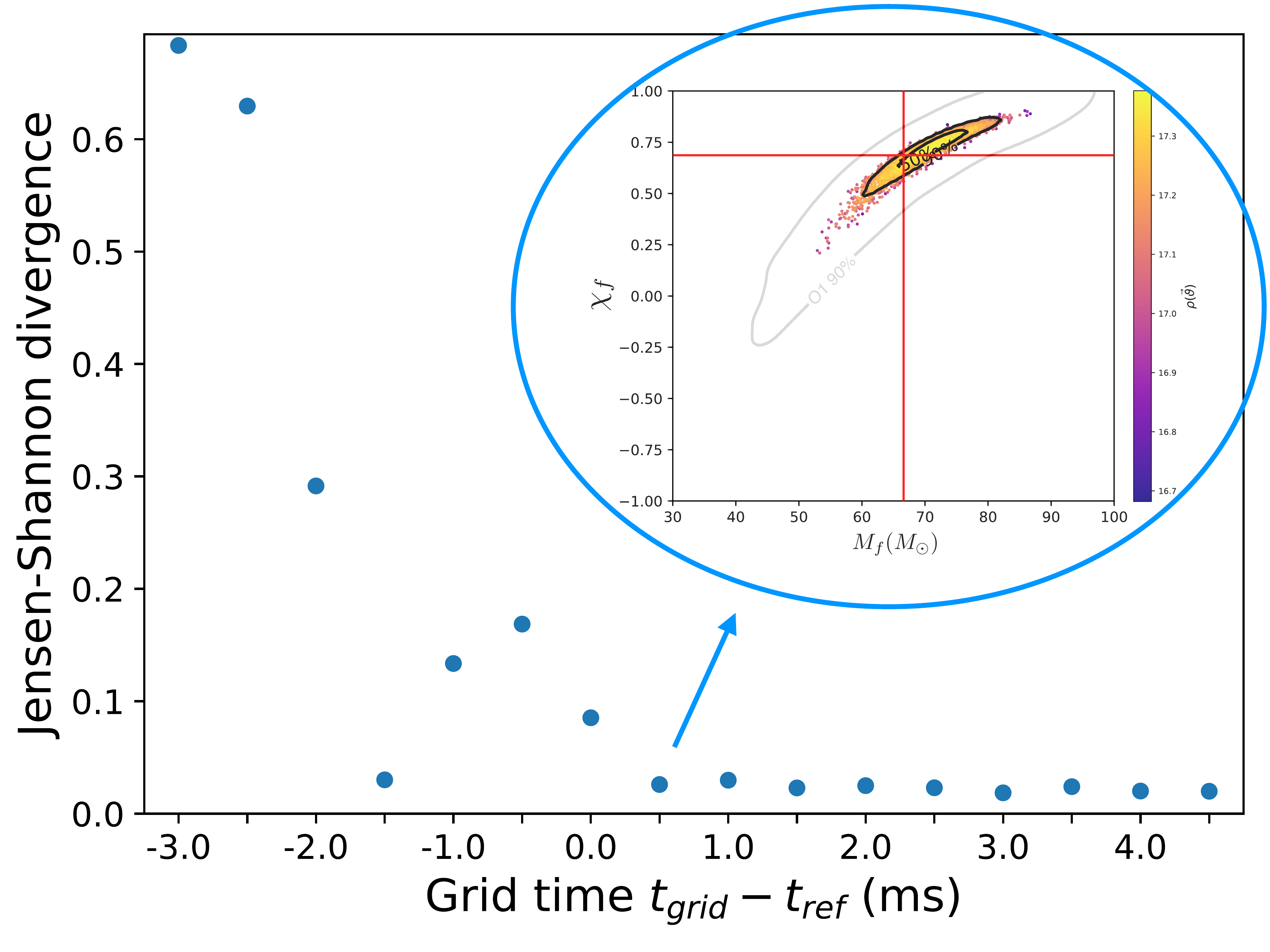}
\caption{
\label{fig:zdhp_jsdiv}
Jensen-Shannon divergence between consecutive grid times
of the inspiral (\textit{top}) and ringdown (\textit{bottom}) analyses
using the ZDHP PSD. Insets show the resulting posterior distribution for
the chosen transition times. The light grey posteriors correspond to the
results obtained with O1 sensitivity. For the inspiral (ringdown) analysis we find that the
JS divergence settles at $\tgrid - \tref=-27.5\,$ ms ($\tgrid - \tref=0.5\,$ ms), consistent with the
O1 PSD results shown in Fig.~\ref{ChooseTime} (Fig.~\ref{RingdownTime}).}
\end{figure}

To ascertain how well we may test the area theorem in the future, we repeat the
entire analysis on the same signal using the zero-detuned high-power (ZDHP) PSD
from Ref. \cite{ZDHP}, which is the design sensitivity for Advanced LIGO.
Figure \ref{fig:zdhp_jsdiv} shows the JS divergence plots for the inspiral
(top) and ringdown analyses (bottom). We find that the JS divergence of the
inspiral (ringdown) posteriors settles to a constant value at $\tref-27.5$ ms ($\tref+0.5$ ms), which is
the same as the O1 results. The selected ringdown grid time corresponds to
$12.6\,M_f$ after the peak amplitude of the signal. While here we have chosen the
JS divergence as an estimate of the adequate ringdown time, there is no unique definition
in the literature of the start of the ringdown~\cite{Thrane:2017lqn,bhagwat2018choosing,PhysRevD.97.044048}.

Results with the ZDHP configuration are shown by the dashed posterior in
Figure~\ref{AreaResult}. In this case, the measured median value of the area
increase with 90\% credible interval is $A_f / A_i = 1.58_{-0.33}^{+0.35}$.
With this sensitivity we obtain a $\sim99.9\%$ probability that the simulated
signal is consistent with the area theorem.

The median value of $A_f / A_i$ underestimates the true value in the O1
results. This is primarily due to a systematic bias arising in the
inspiral analysis. Both the mass ratio $q \equiv M_1 / M_2 \, (M_1 \geq M_2)$ and
total mass are overestimated with this method; this leads to an overestimate
of $A_i$ by $\sim10\%$, in turn leading to an underestimate of $A_f / A_i$.
Further biasing the median is the fact that the simulated signal used here has
$q=1$. Since this is on the boundary of allowed parameter space, the median of
the posterior distribution can only ever overestimate $q$, again leading to an
underestimate of $A_f / A_i$. The chirp mass is well measured, therefore an overestimate
in the mass ratio leads to an overestimate in the total mass as well.
 Finally, as seen in Figure~\ref{AreaResult},
our prior on $A_f / A_i$ strongly favours a violation of the area theorem. The
prior follows from the uniform prior on the inspiral masses and spins and the
uniform prior on the ringdown frequency and damping time. This also shifts the
posterior distribution to smaller values of $A_f / A_i$, though the effect is
small compared to the effect of the systematic bias.

Overall, the systematic bias in the area increase is less than the statistical
error. Furthermore, as the bias is toward violations of the area theorem, it
is a conservative error when evaluating the credible interval to which the
signal is consistent with the area theorem.

Given the measured initial parameters from the inspiral analysis,
one can obtain the expected area change for each point in the 
initial distribution using the fitting formulae
in~\cite{PhysRevD.93.044006}. A direct comparison of the expected change
with the measured change indicates the level of agreement of the
final object with the fitting formulae, and therefore with general relativity. 
If the final object agrees with general relativity,
the ratio between the measured and the expected values should be 1.
The bottom panel in Figure~\ref{AreaResult} shows this ratio for the
area change, $\Delta A = A_f - A_i$.

\begin{figure}[tb]
\centering
\includegraphics[width=\columnwidth]{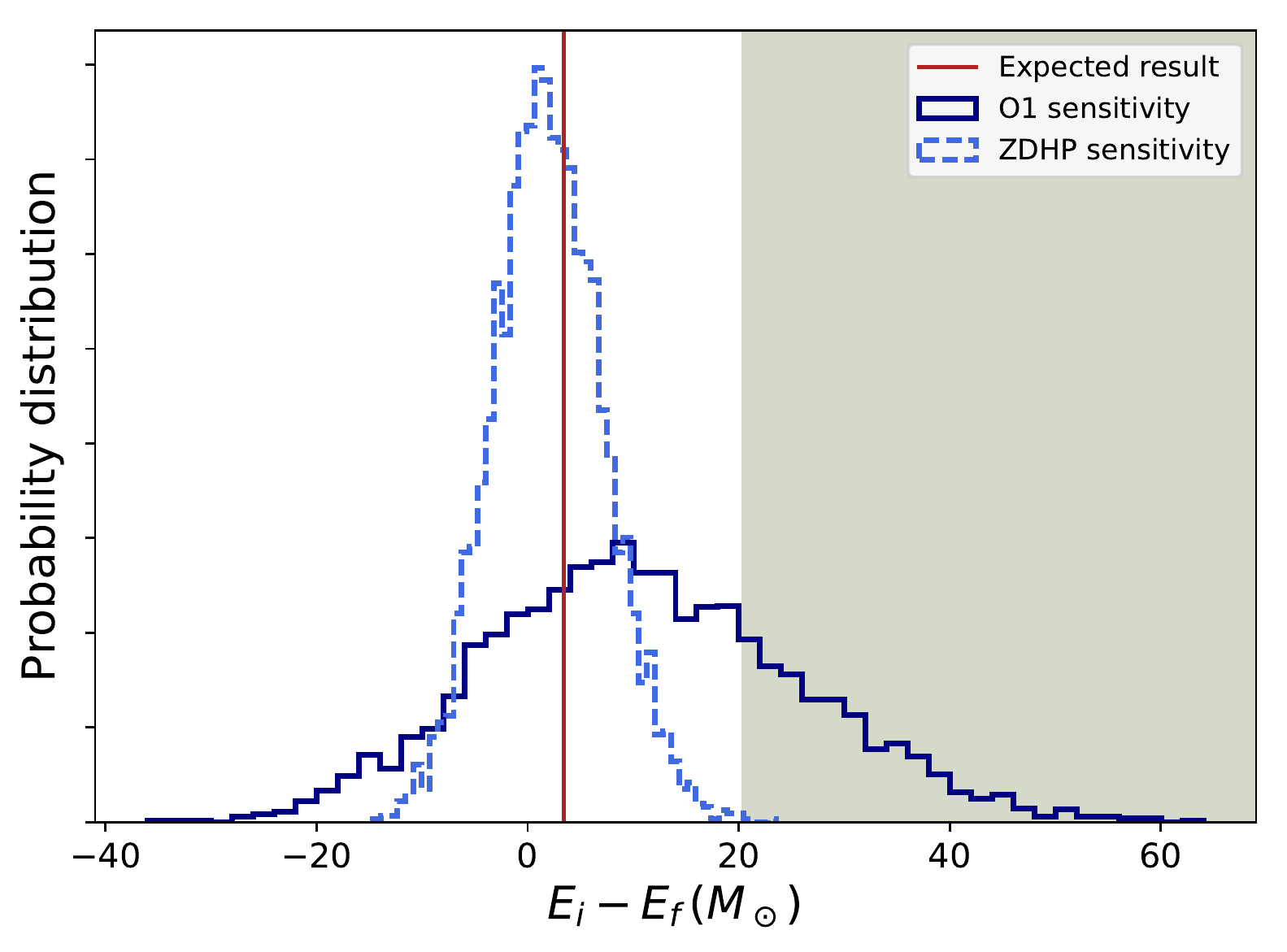}
\includegraphics[width=\columnwidth]{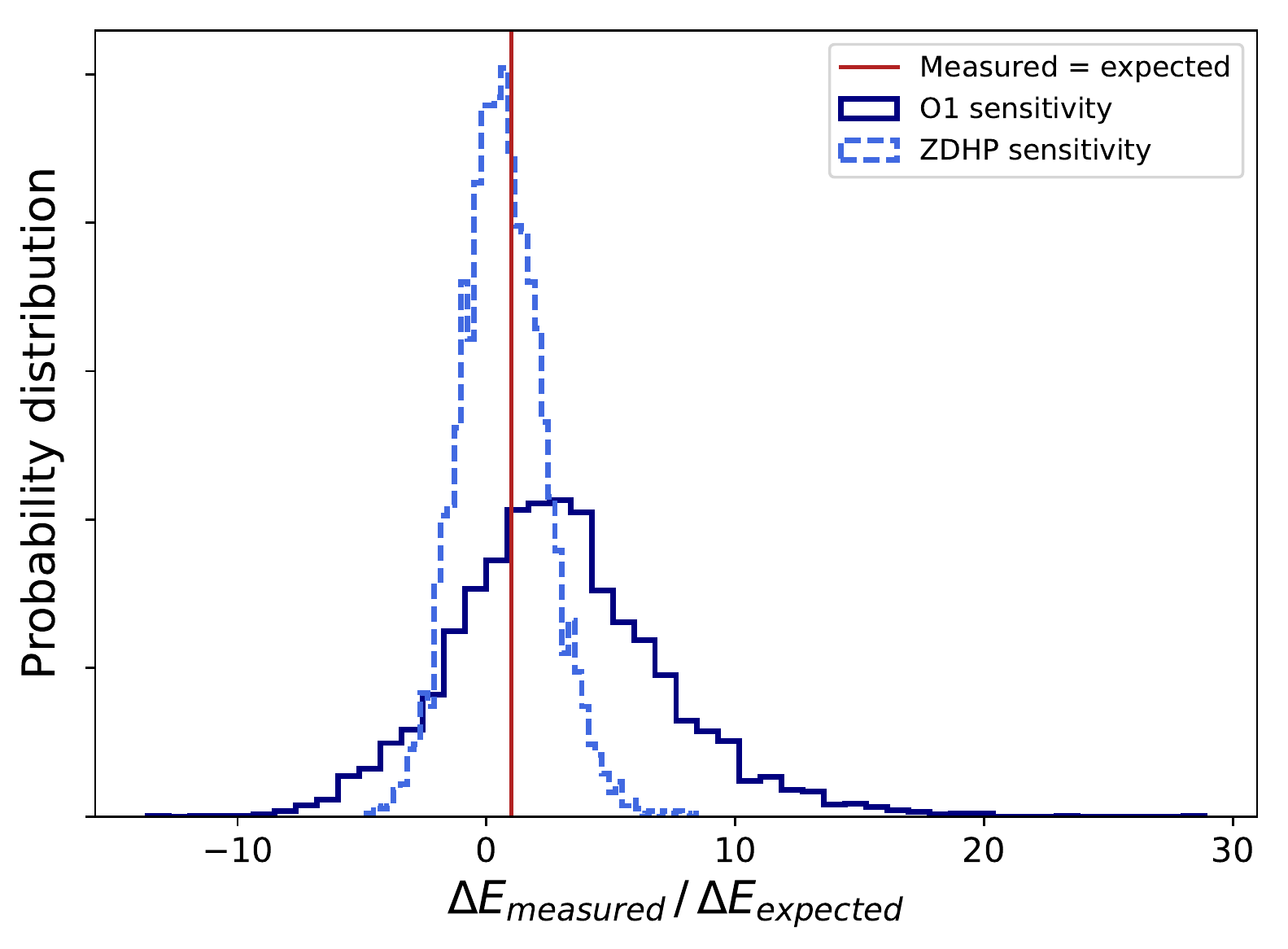}
\caption{\label{EnergyResult} (\textit{Top}) Posterior distribution on the energy radiated during the coalescence. The red line indicates the expected value for the injected parameters, $E_i - E_f \simeq 3.4 M_\odot$. The shaded region shows the theoretical limit of 29\% in the energy emitted. (\textit{Bottom}) Distribution of the ratio $\Delta E_{\text{measured}} / \Delta E_{\text{expected}}$. The expected radiated energy is given by the initial parameters obtained in the inspiral analysis and the corresponding expected final parameters from the fitting formulae in~\cite{PhysRevD.93.044006}. The vertical red line indicates agreement between the measured and the expected values, i.e. $\Delta E_{\text{measured}} / \Delta E_{\text{expected}} = 1$.}
\end{figure}

The independent measurements of the initial and final mass performed in this
work also allow for estimating the energy radiated away by the system.
The top panel in Figure~\ref{EnergyResult} shows the posterior distribution in the energy
radiated away, $\Delta E = E_i - E_f$, for both current and future Advanced LIGO
sensitivities. The shaded region indicates the 29\% bound derived from the area
increase law. Similar as with the area change, one can compare the measured
energy radiated, $\Delta E_{\text{measured}}$, with the result one would
obtain making use of the fitting formulae in~\cite{PhysRevD.93.044006}, $\Delta E_{\text{expected}}$.
The bottom panel in Figure~\ref{EnergyResult} shows the ratio between these two energies.

\begin{figure}[tb]
\centering
\includegraphics[width=\columnwidth]{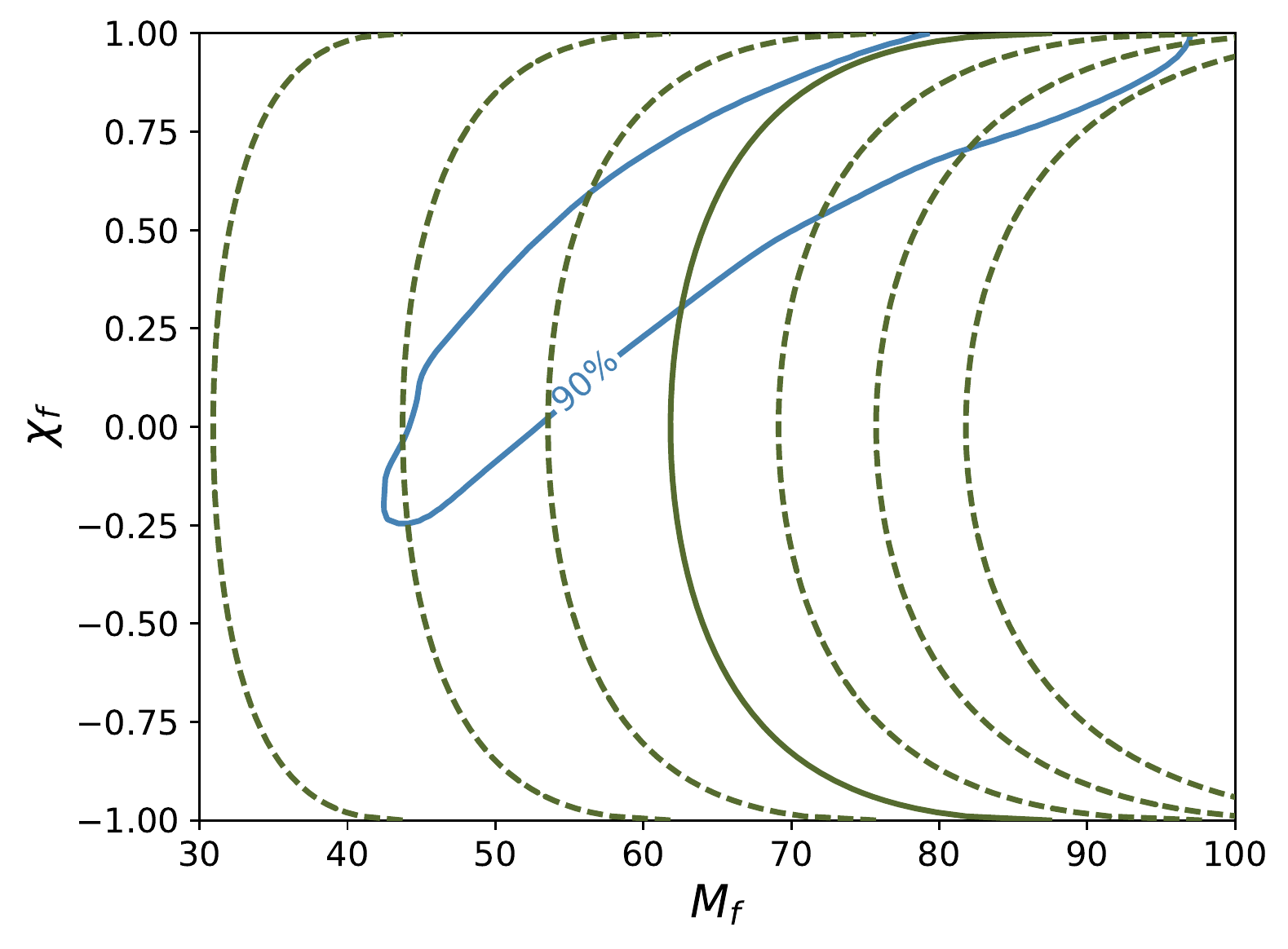}
\caption{\label{ConstantArea} Lines of constant area as function of the final mass and spin. The solid line shows the expected value, while dashed lines indicate areas $\pm 25\%, 50\%, 75\%$. For comparison, the 90\% credible interval from the ringdown analysis with O1 sensitivity is shown.}
\end{figure}

As can be seen by comparison of Figure~\ref{AreaResult} and Figure~\ref{EnergyResult},
the measurement of the area increase is more accurate than the 
measurement of the energy radiated. This can be understood from the
lines of constant area shown in Figure~\ref{ConstantArea}. The area follows the
ringdown posteriors better than the mass at positive spins, which are expected
for two initially non-spinning black holes. Furthermore, we can expect the measurement
of the area to be even sharper for highly aligned spinning black holes.%, while the spin
%does not significantly affect the measurement of the mass.

\section{Simulating violations of the area theorem}
\label{sec:violations}

In this Section we explore if we would be able to measure a violation
of the area increase law with the method described above.  We do not
explore here how this violation could happen, but only if we would be
able to measure a violation.  For this purpose we compare our inspiral
measurements with lower-mass ringdown signals. The spin of the final
object is only dependent on the mass ratio and the spins of the initial
objects. Changing the total mass of the binary will yield different
final mass but the same final spin. Therefore, we find that for our
system, a violation of the area theorem would require $M_f < 53.3
M_\odot$. We use two ringdowns with masses $M_f \simeq 52.3 M_\odot$
and $M_f \simeq 47.6 M_\odot$, and perform the ringdown analysis only
with the ZDHP sensitivity. 

\begin{figure}[tb]
\centering
\includegraphics[width=\columnwidth]{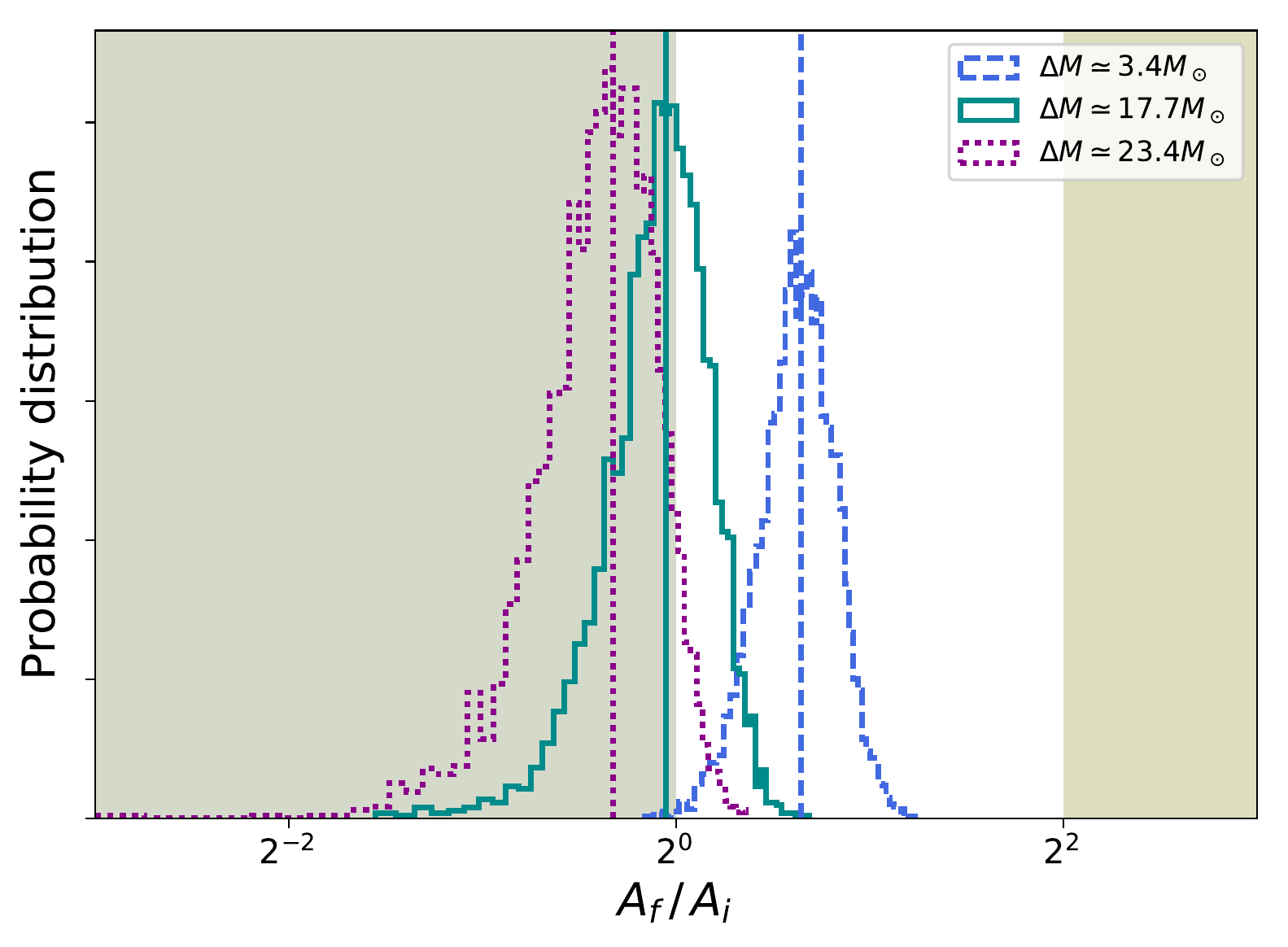}
\caption{\label{AreaViolation} Posterior distribution on the ratio of the final to initial areas, $A_f / (A_1 + A_2)$, with ZDHP sensitivity. The shaded region $A_f/A_i < 1$ indicates violation of the area theorem. The shaded region $A_f / A_i > 4$ indicates violation of the conservation of energy. Vertical lines indicate the expected value for each case. The dashed posterior distribution is the result shown in the previous section. The solid posterior distribution is a signal that slightly violates the area theorem, and the dotted posterior distribution is a signal that clearly violates the area theorem.}
\end{figure}

Figure~\ref{AreaViolation} shows the result of combining these new 
ringdowns with the original inspiral results. The dashed posterior distribution with 
$\Delta M = M_f - M_t \simeq 3.4 M_\odot$ is the result 
shown in the previous section. The continuous line with
$\Delta_M \simeq 17.7 M_\odot$ is the system that yields a small violation
of the area theorem. We find a $\sim55.2\%$ probability that this system 
is in violation of the area theorem. The dotted line shows the system with 
even lower mass, for which we obtain $\sim91.1\%$ probability that 
the system is in violation of the area theorem.

\section{Conclusions}
\label{sec:conclusion} 

We developed a method to test the area theorem on gravitational-wave signals
from binary black hole coalescences. This method completely ignores information
from the highly dynamical merger phase, thus ensuring that the initial and
final parameters are measured independently from each other, and without
assuming general relativity during the merger process.

With current gravitational-wave detectors sensitivities and zero noise, 
we obtained $\sim 74.6\%$ probability that a numerical
waveform similar to GW150914 is consistent with the area theorem. 
This probability increases to $\sim99.9\%$ with design 
sensitivity (ZDHP) for Advanced LIGO.

The next obvious step is to perform this analysis on a real
gravitational-wave signal.  From the binary black holes known to date
there is only one with loud enough ringdown to perform this test:
GW150914. We plan to show results on this event in a future
publication.  One of the major issues is the choice of the
start time for the ringdown.  Our method of using the
JS divergence will be a useful starting point for this analysis but
might need to be refined for real data. In simulations with Gaussian
noise, we have found that the qualitative behaviour of the JS
divergence is the same as in Fig.~\ref{RingdownTime}, however the data
has larger variations.  This will be investigated in future work
on real detector data.

The confidences obtained are mainly bound by the ringdown analysis, which
yields wider posterior distributions than the inspiral analysis. Higher modes
could start becoming important before the end of the second generation of
gravitational-wave detectors. The addition of sub-dominant modes to the
ringdown analysis will not only allow for tests of the Kerr nature of the final black
holes, but could result in better constraints on the final mass and spin.
Therefore, detectable higher modes could improve the confidence level on
testing the area theorem.

As we have seen, the sky location is also an important limiting factor in our
analysis. In this paper we have shown results on a two-detector network.
However, the Virgo detector joined the second generation of gravitational-wave
detectors in August of 2017, and showed an important contribution on sky
localisation of gravitational-wave sources. Future events could therefore show
different confidence levels as reported in this work if more than two detectors
are being used.

The techniques we present may also be extended to make a joint measurement of
the violation of the area theorem for a population of binary black hole mergers. As
an increasing number of mergers are detected, we expect that this may provide
significant improvements to the overall uncertainties.

\acknowledgments

We would like to thank Juan Calder\'{o}n-Bustillo, Thomas Dent, 
Ben Farr, Will Farr, Domenico Giulini, David Keitel, Sebastian Khan, 
Andrew Lundgren, Frank Ohme and Reinhard Prix for interesting discussions.
This research has made use of data, software and/or web tools obtained
from the LIGO Open Science Centre (https://losc.ligo.org), a service
of LIGO Laboratory and the LIGO Scientific Collaboration. LIGO is
funded by the U.S. National Science Foundation.

\bibliography{areatheorem}{}
\end{document}